\documentclass[12pt]{article}

\usepackage{amsmath,amssymb,amsfonts,amsbsy}
\usepackage{cite}
\usepackage{graphicx}
\usepackage{wrapfig}
\usepackage[font=small,labelfont=bf,labelsep=period]{caption}

  \newcommand{\lsim}{\raisebox{-4pt}{$\,\stackrel{\textstyle
							   <}{\sim}\,$}}
  \newcommand{\gsim}{\raisebox{-4pt}{$\,\stackrel{\textstyle
							   >}{\sim}\,$}}
  
  \newcommand{\be}{\begin{equation}}
  \newcommand{\ee}{\end{equation}}
  \newcommand{\ba}{\begin{eqnarray}}
  \newcommand{\ea}{\end{eqnarray}}
  \newcommand{\req}[1]{(\ref{#1})}
  \def\={\,=\,}
  \newcommand{\ci}[1]{\cite{#1}}

  \def\gev{~{\rm GeV}}

  \newcommand{\tw}{\textwidth}

  \def\vb0{{\bf b}_0}

  \def\={\,=\,}

\begin{document}
\addcontentsline{toc}{subsection}{{Transversity in exclusive meson electroproduction}\\
{\it B.B. Kroll}}

%%%%%%% please do not touch these! %%%%%%
\setcounter{section}{0}
\setcounter{subsection}{0}
\setcounter{equation}{0}
\setcounter{figure}{0}
\setcounter{footnote}{0}
\setcounter{table}{0}

\begin{center}
\textbf{TRANSVERSITY  IN EXCLUSIVE MESON ELECTROPRODUCTION}   

\vspace{5mm}

P.~Kroll %$^{\,\dag}$

\vspace{5mm}

\begin{small}
  \emph{ Fachbereich Physik, Universit\"at Wuppertal, D-42097 Wuppertal,  Germany} \\
  %$\dag$ 
   \emph{E-mail: kroll@physik.uni-wuppertal.de}
\end{small}
\end{center}

\vspace{0.0mm} % Don't laugh: it does change the spacing!

\begin{abstract}
In this talk various spin effects in hard exclusive electroproduction of mesons
 are briefly reviewed and the data discussed in the light of recent
 theoretical calculations within the frame work of the handbag
 approach. For $\pi^+$ electroproduction it is shown that there is a
 strong contribution from $\gamma^*_T\to \pi$ transitions which can be
 modeled by the transversity GPD $H_T$ accompanied by the twist-3
 meson wave function.
 %This talk has been presented at the Spin Conference held at Dubna, September, 2009.
\end{abstract}

\vspace{7.2mm} 
%%%%%%%%%%%%%%%%%%%%%%%%%%%%%%%%%%%%%%%%%%%%%%%%%%%%%%%%%%%%%%%%%%%%%%%%%%%%%%%
  \section{Introduction}
  \label{sec:intro}
\begin{wrapfigure}[15]{L}{0.48\tw}
\centering
\vspace*{-3mm}
\includegraphics[width=0.48\tw,bb=103 462 388
658,clip=true]{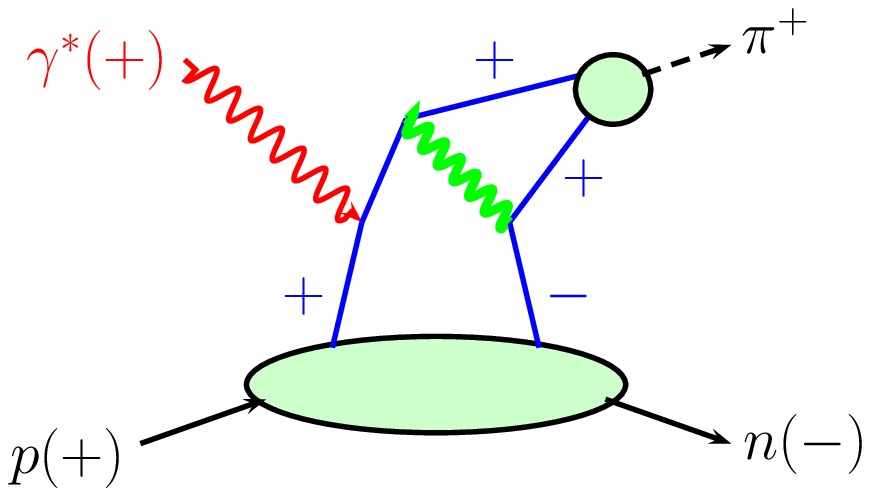}

\vspace*{-12mm}
\caption{\label{fig:1} A typical lowest order Feynman graph for meson electroproduction.
The signs indicate helicity labels for the contribution from transversity GPDs
to the amplitude ${\cal M}_{0-,++}$, see text.}
\end{wrapfigure}

%%%%%%%%%%%%%%%%%%%%%%%%%%%%%%%%%%%%%%%%%%%%%%%%%%%%%%%%%%%%%%%%%%%%%%%%%%%
Electroproduction of mesons allows for the measurement of many spin
effects. For instance, using a longitudinally or transversely polarized target
and/or a longitudinally polarized beam various spin asymmetries can be
measured. The investigation of spin-dependent observables allows for a deep
insight in the underlying dynamics. Here, in this article, it will be reported 
upon some spin effects and their dynamical interpretation in the frame work of 
the so-called handbag approach which offers a partonic description of meson 
electroproduction provided the virtuality of the exchanged photon, $Q^2$, is 
sufficiently large. The theoretical basis of the handbag approach is the 
factorization of the process amplitude into a hard partonic subprocess and in 
soft hadronic matrix elements, the so-called generalized parton distributions 
(GPDs), as well as wave functions for the produced mesons, see Fig.\
\ref{fig:1}. In collinear approximation factorization has been shown to hold 
rigorously for hard exclusive meson electroproduction~\ci{rad96,col96}. It has 
also been shown that the transitions from a longitudinally polarized photon to 
a likewise polarized vector meson or a pseudoscalar one, $\gamma^*_L\to V_L(P)$, 
dominates at large $Q^2$. Other photon-meson transitions are suppressed by
inverse powers of the hard scale.

Here, in this article a variant of the handbag approach is utilized for the
interpretation of the data in which the subprocess amplitudes are calculated 
within the modified perturbative approach~\ci{sterman}, and the GPDs are
constructed from reggeized double distributions~\ci{mul94,rad98}. In the 
modified perturbative approach the quark transverse momenta are retained in 
the subprocess and Sudakov suppressions are taken into account. The partons 
are still emitted and re-absorbed by the proton collinearly. For the meson 
wave functions Gaussians in the variable $k_\perp^2/(\tau(1-\tau))$ are 
assumed with transverse size parameters fitted to experiment~\ci{jakob}. The 
variable $\tau$ denotes the fraction of the meson's momentum the quark
entering the meson, carries. In a series of papers~\ci{GK1} it has been shown
that with the proposed handbag approach the data on the cross sections and 
spin density matrix elements (SDMEs) for $\rho^0$ and $\phi$ production are 
well fitted in the kinematical range of $Q^2\gsim 3\,\gev^2$, 
$\;W\gsim 5\,\gev\;$ (i.e.\ for small values of skewness $\;\xi\simeq x_{Bj}
/2\,\lsim\, 0.1\;$) and 

\noindent for the squared invariant momentum transfer $-t^\prime = -t+t_0\, 
\lsim\, 0.6\,\gev^2$ where $t_0$ is the value of $t$ for forward
scattering. This analysis fixes the GPD $H$ for quarks and gluons
quite well. The other GPDs do practically not contribute to the 
cross sections and SDMEs at small skewness.

As mentioned spin effects in hard exclusive meson electroproduction will be
briefly reviewed and their implications on the handbag approach and above all
for the determination of the GPDs, discussed. In Sect.\ 2 the role of target 
spin asymmetries in meson electroproduction is examined. Sect.\ 3 is devoted 
to a discussion of the the target spin asymmetries in pion electroproduction 
and Sect.\ 4 to those for vector mesons and the GPD $E$. Finally, in Sect.\ 5, 
a summary is presented. 

%%%%%%%%%%%%%%%%%%%%%%%%%%%%%%%%%%%%%%%%%%%%%%%%%%%%%%%%%%%%%%%%%%%%%%%%%%%%%%%%%%%%%%%%%%%%%
\section{Target asymmetries}
%%%%%%%%%%%%%%%%%%%%%%%%%%%%%%%%%%%%%%%%%%%%%%%%%%%%%%%%%%%%%%%%%%%%%%%%%%%%%%%%%%%%%%%%%%%%
The electroproduction cross sections measured with a transversely or
longitudinally polarized target consist of many terms, each can be projected
out by a $\sin{\varphi}$ or $\cos{\varphi}$ moment where $\varphi$ is a linear 
combination of $\phi$, the azimuthal angle between the lepton and the hadron 
plane and $\phi_s$, the orientation of the target spin vector~\ci{sapeta}. In
Tab.\ \ref{tab:1} the features of some of these moments are displayed. As the 
dominant interference terms reveal the target asymmetries provide detailed 
information on the $\gamma^* p\to MB$ amplitudes and therefore on the
underlying dynamics that generates them.    

\begin{figure}[t]
  \begin{center}
  \includegraphics[width=0.44\tw,bb=25 333 532 743,clip=true]{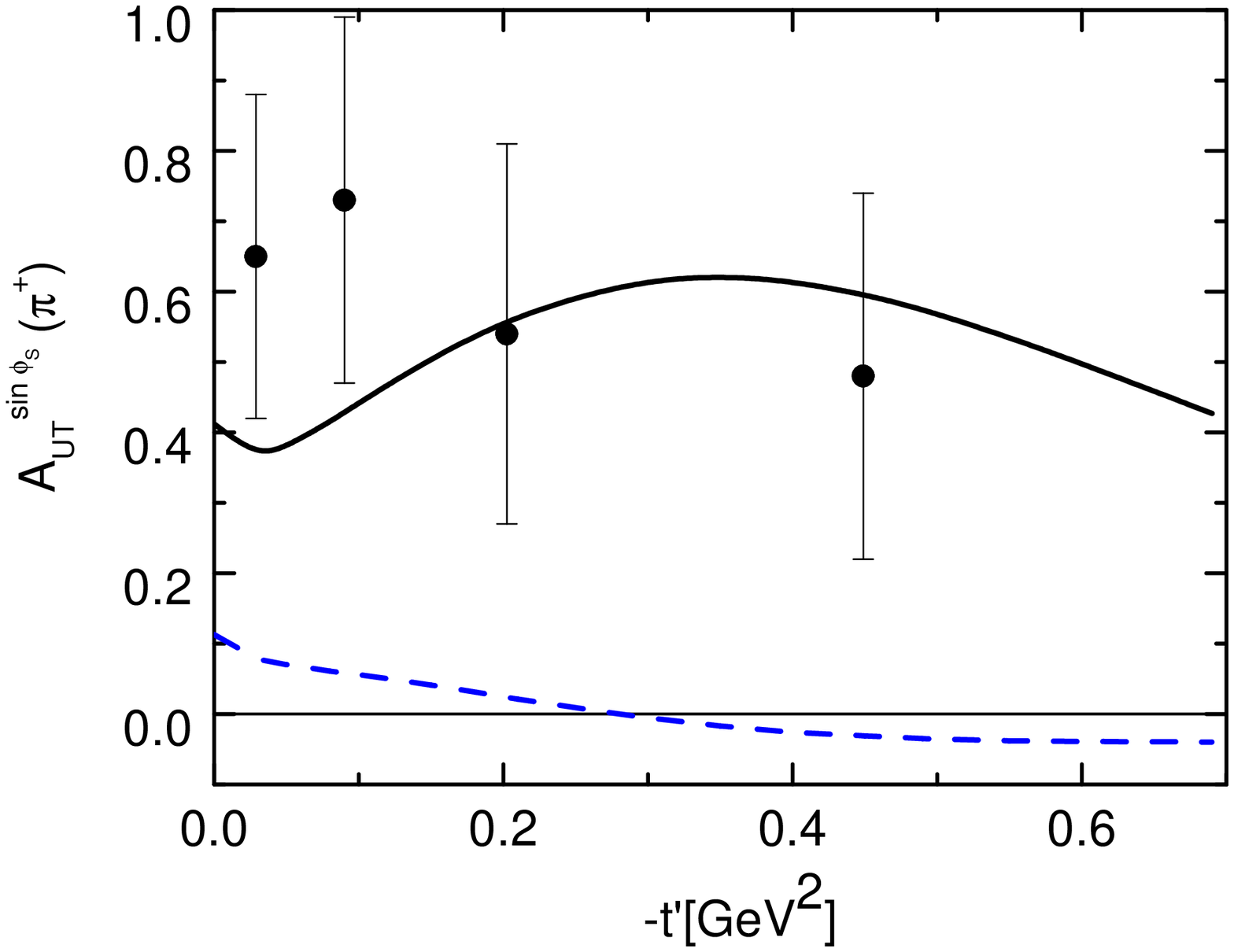}
\includegraphics[width=0.41\tw , bb= 44 327 533 753,clip=true]{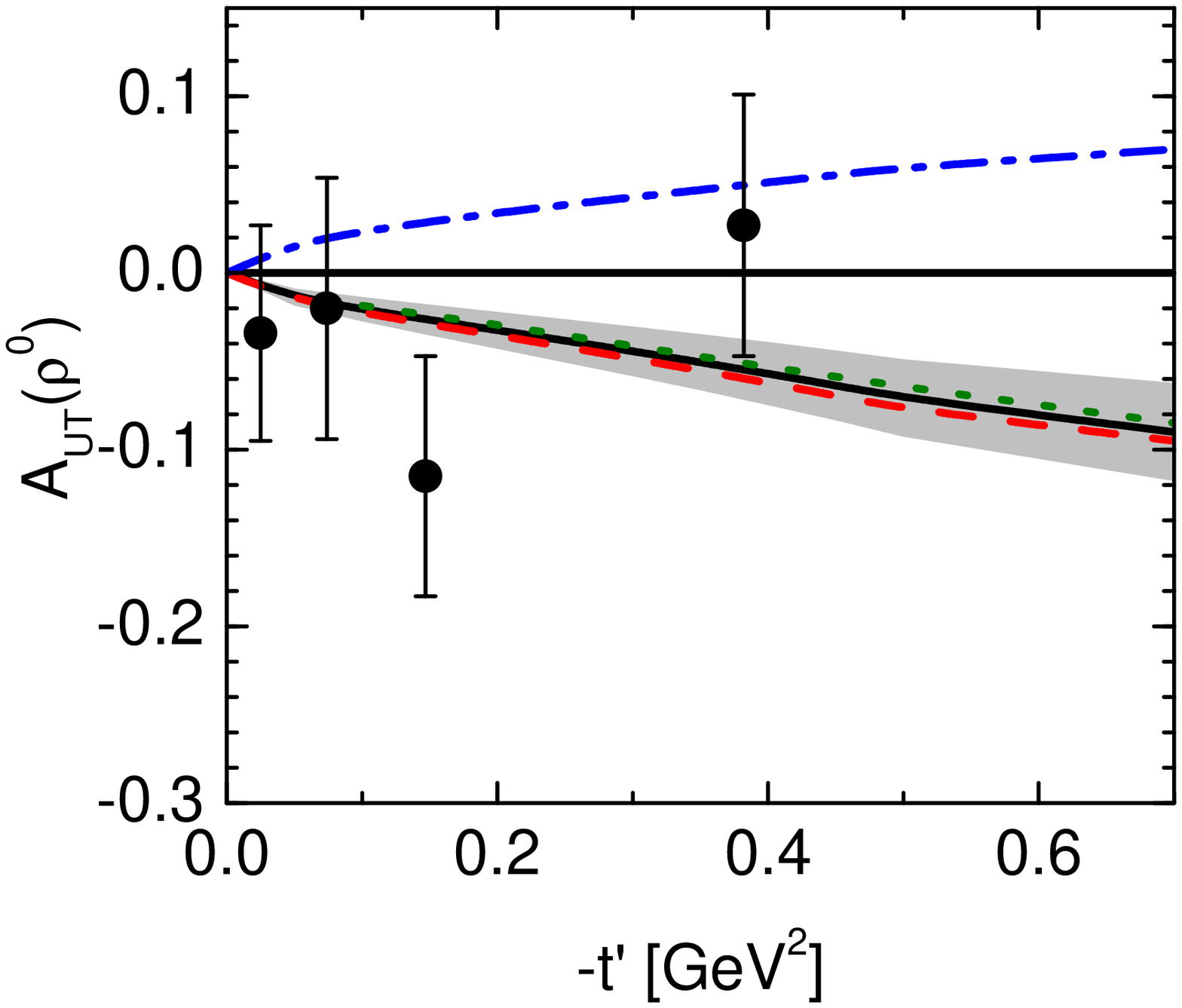}
  \caption{\label{fig:2} The $\sin{\phi_s}$ moment for a
   transversely polarized target at $Q^2\simeq 2.45\,\gev^2$ and
    $W=3.99\,\gev$ for $\pi^+$ production. The predictions from the handbag
    approach of Ref.\ \protect\ci{GK5} are shown as a solid line. The 
    dashed line is obtained disregarding the twist-3 contribution. Data are 
    taken from Ref.\ \protect\ci{Hristova}.}
\caption{\label{fig:3} The asymmetry $A_{UT}^{\sin{(\phi-\phi_s)}}$ for $\rho^0$ 
    production at $W=5\,\gev$ and $Q^2=2\,\gev^2$. Data taken from Ref.\ 
    \protect\ci{HERMES-rho}. The lines represent the results presented 
    in Ref.\ \protect\ci{GK4}. For further notations see text and Ref.\ 
    \protect\ci{GK4}.} 
\end{center}
\end{figure}

A number of these moments have been measured recently. A particularly striking
result is the $\sin{\phi_S}$ moment which has been measured by the HERMES 
collaboration for $\pi^+$ electroproduction~\ci{Hristova}. The data on this
moment, shown in Fig.\ \ref{fig:2}, exhibit a mild $t$-dependence and do not 
show any indication for a turnover towards zero for $t^\prime\to 0$. Inspection 
of Tab.\ \ref{tab:1} reveals that this behavior of $A_{UT}^{\sin{\phi_s}}$ at 
small $-t^\prime$ requires a contribution from the interference term 
${\rm Im}\big[{\cal M}_{0-,++}^*\,{\cal M}_{0+,0+}\big]$. Both the contributing 
amplitudes are helicity non-flip ones and are therefore not forced to vanish 
in the forward direction by angular momentum conservation. Thus, we see that 
for pion electroproduction there are strong contributions  from 
$\gamma^*_T \to \pi$ transitions. The underlying dynamical mechanism for such
transitions will be discussed in Sect.\ 3.

\begin{table*}[t]
  \renewcommand{\arraystretch}{1.4}
  \centering
  {\begin{tabular}{|c|| c | c | c | c |}
  \hline     
   observable  & dominant &   amplitudes  & low $t^\prime$ \\
	  & interf. term  &   &  behavior \\[0.2em]   
  \hline
  $A_{UT}^{\sin(\phi-\phi_s)}$ &  LL  & ${\rm Im}\big[{\cal M}^*_{0-,0+}
			 {\cal M}_{0+,0+}\big]$ & $\propto \sqrt{-t^\prime}$   \\[0.2em]
  $A_{UT}^{\sin(\phi_s)}$ & LT  &  ${\rm Im}\big[{\cal M}^*_{0-,++}{\cal M}_{0+,0+}\big]$  & const.  \\[0.2em]
  $A_{UT}^{\sin(2\phi-\phi_s)}$ & LT & ${\rm Im}\big[{\cal M}^*_{0\mp,-+}
			       {\cal M}_{0\pm,0+}\big]$ &  $\propto t^\prime$\\[0.2em]
  $A_{UT}^{\sin(\phi+\phi_s)}$ & TT &  ${\rm Im}\big[{\cal M}^*_{0-,++}
			  {\cal M}_{0+,++}\big]$ & $\propto \sqrt{-t^\prime}$    \\[0.2em]
  $A_{UT}^{\sin(2\phi+\phi_s)}$ & TT &  $\propto \sin{\theta_\gamma}$ &  $\propto t^\prime$\\[0.2em]
  $A_{UT}^{\sin(3\phi-\phi_s)}$ & TT & ${\rm Im}\big[{\cal M}^*_{0-,-+}
			       {\cal M}_{0+,-+}\big]$ & $\propto (-t^\prime)^{(3/2)}$  \\[0.2em]
  \hline
  $A_{UL}^{\sin(\phi)}$   & LT & ${\rm Im}\big[{\cal M}^*_{0-,++} {\cal M}_{0-,0+}\big]$ & 
  $\propto \sqrt{-t^\prime}$   \\[0.2em]
  \hline
  \end{tabular}
\caption{Features of the asymmetries for transversally and longitudinally
    polarized targets. The angle $\theta_\gamma$ describes the rotation in the 
    lepton plane from the direction of the incoming lepton to the virtual
    photon one; it is very small.}
    %\end{center}
  \label{tab:1}}
 \renewcommand{\arraystretch}{1.0}   
  \end{table*}

For $\rho^0$ production the $\sin{(\phi-\phi_s)}$ moment has been measured by
HERMES~\ci{HERMES-rho} and COMPASS~\ci{COMPASS-rho}; 
the latter data being still preliminary. The HERMES data are shown in Fig.\ 
\ref{fig:3}. In the handbag approach $A_{UT}^{\sin{(\phi-\phi_s)}}$ can also
be expressed by an interference term of the convolutions of the GPDs $H$
and $E$ with hard scattering kernels 
\be
A_{UT}^{\sin{\phi-\phi_s}} \sim {\rm Im} \langle E \rangle^* \langle H \rangle
\ee
instead of the helicity amplitudes. Given that $H$ is known from the analysis
of the $\rho^0$ and $\phi$ cross sections and SDMEs, $A_{UT}$ provides 
information on $E$~\ci{GK4}.  In order to calculate this target asymmetry $E$
is needed. What is known about the GPD $E$ will be recapitulated in
Sect.\ 4.

%%%%%%%%%%%%%%%%%%%%%%%%%%%%%%%%%%%%%%%%%%%%%%%%%%%%%%%%%%%%%%%%%%%%%%%%%%%%%%%%%%%%%
\section{Target spin asymmetries in $\pi^+$ production}
%%%%%%%%%%%%%%%%%%%%%%%%%%%%%%%%%%%%%%%%%%%%%%%%%%%%%%%%%%%%%%%%%%%%%%%%%%%%%%%%%%%%%
In Ref.\ \ci{GK5} electroproduction of positively charged pions has been
investigated in the same handbag approach as applied to vector meson 
production~\ci{GK1}. To the asymptotically leading amplitudes for 
longitudinally polarized photons the GPDs $\widetilde{H}$ and $\widetilde{E}$ 
contribute in the isovector combination 
\be
\widetilde{F}^{(3)}=\widetilde{F}^u_v-\widetilde{F}^d_v\,.
\ee
instead of $H$ and $E$ for vector mesons. In deviation to work performed in 
collinear approximation the full electromagnetic form factor of the pion as 
measured by the $F_\pi-2$ collaboration~\ci{horn06} is naturally 
taken into account~\footnote
{As compared to other work $\widetilde{E}$ contains only the non-pole contribution.}
(see also the recent work by Bechler and Mueller~\ci{bechler}). 
The GPDs $\widetilde{H}$ and $\widetilde{E}$ are again constructed with the 
help of double distributions with the forward limit of $\widetilde{H}$ being 
the polarized parton distributions while that of $\widetilde{E}$ is 
parameterized analogously to the familiar parton distributions
\be
\tilde{e}^u = - \tilde{e}^d = \widetilde{N}_e x^{-0.48} (1-x)^5\,,
\ee
with $\widetilde{N}_e$ fitted to experiment.

As is mentioned in Sect.\ 2 experiment requires a strong contribution from the
helicity-non-flip amplitude ${\cal M}_{0-,++}$ which does not vanish in the
forward direction. How can this amplitude be modeled in the frame work of the 
handbag approach? From the usual helicity non-flip GPDs $H, E, \ldots$ one
obtains a contribution to ${\cal M}_{0-,++}$ that vanishes $\propto t^\prime$
if it is non-zero at all. However, there is a second set of GPDs, the 
helicity-flip or transversity ones $H_T, E_T, \ldots$~\ci{diehl01,hoodbhoy}. As 
inspection of Fig.\ \ref{fig:1} where the helicity configuration of the
process is specified, reveals the proton-parton vertex is of non-flip nature
in this case and, hence, is not forced to vanish in the forward direction by 
angular momentum conservation. One also sees from Fig.\ \ref{fig:1}, that the 
helicity configuration of the subprocess is the same as for the full amplitude. 
Therefore, also the subprocess amplitude has not to vanish in the forward 
direction and so the full amplitude. The prize to pay is that quark and 
antiquark forming the pion have the same helicity. Therefore, the twist-3 pion 
wave function is needed instead of the familiar twist-2 one. The dynamical 
mechanism building up the amplitude ${\cal M}_{0-,++}$ is so of twist-3 order. 
This mechanism has been first proposed in Ref.\ \ci{passek} 
for photo- and electroproduction of mesons where $-t$ is considered as the 
large scale~\ci{huang}.

In Ref.\ \ci{GK5} the twist-3 pion wave function is taken from Ref.\ 
\ci{braun90} with the three-particle Fock component neglected. This wave 
function, still containing a pseudoscalar and a tensor component, is
proportional to the parameter $\mu_\pi=m^2_\pi/(m_u+m_d) \simeq 2\,\gev$ 
at the scale of $2\,\gev$ as a consequence of the divergency of the
axial-vector current ($m_u$ and $m_d$ are current quark masses). It is 
further assumed that the dominant transversity GPD is $H_T$ while the other
three can be neglected. The forward limit of $H^a_T$ is the transversity 
distribution $\delta^a(x)$ which has been determined in \ci{anselmino} 
in an analysis of data on the asymmetries in semi-inclusive electroproduction
of charged pions measured with a transversely polarized target. Using these
results for $\delta^a(x)$  the GPDs $H_T^a$ have been modeled in a manner 
analogously to that of the other GPDs ( see Eq.\ \req{E-ansatz})~\footnote
{While the relative signs of $\delta^u$ and $\delta^d$ is fixed in the
  analysis performed in Ref.\ \ci{anselmino} the absolute sign is not. 
  Here, in $\pi^+$ electroproduction a positive $\delta^u$ which goes along
  with a negative $\delta^d$ is required by the signs of the target asymmetries.}.

\begin{figure}[t]
  \begin{center}
  \includegraphics[width=0.44\tw,bb=16 349 533 743,clip=true]{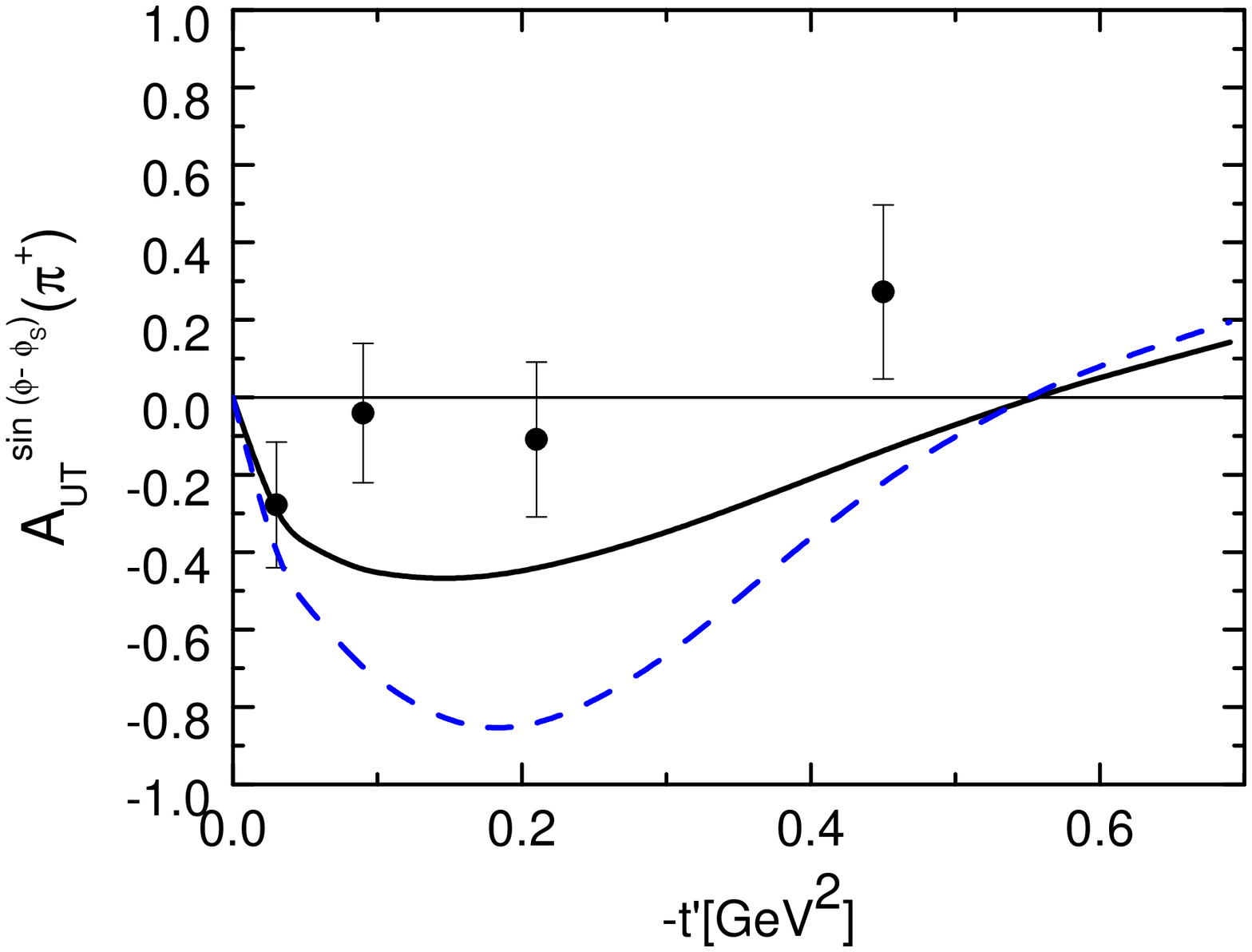}
  \includegraphics[width=0.42\tw,bb=30 346 533 746,clip=true]{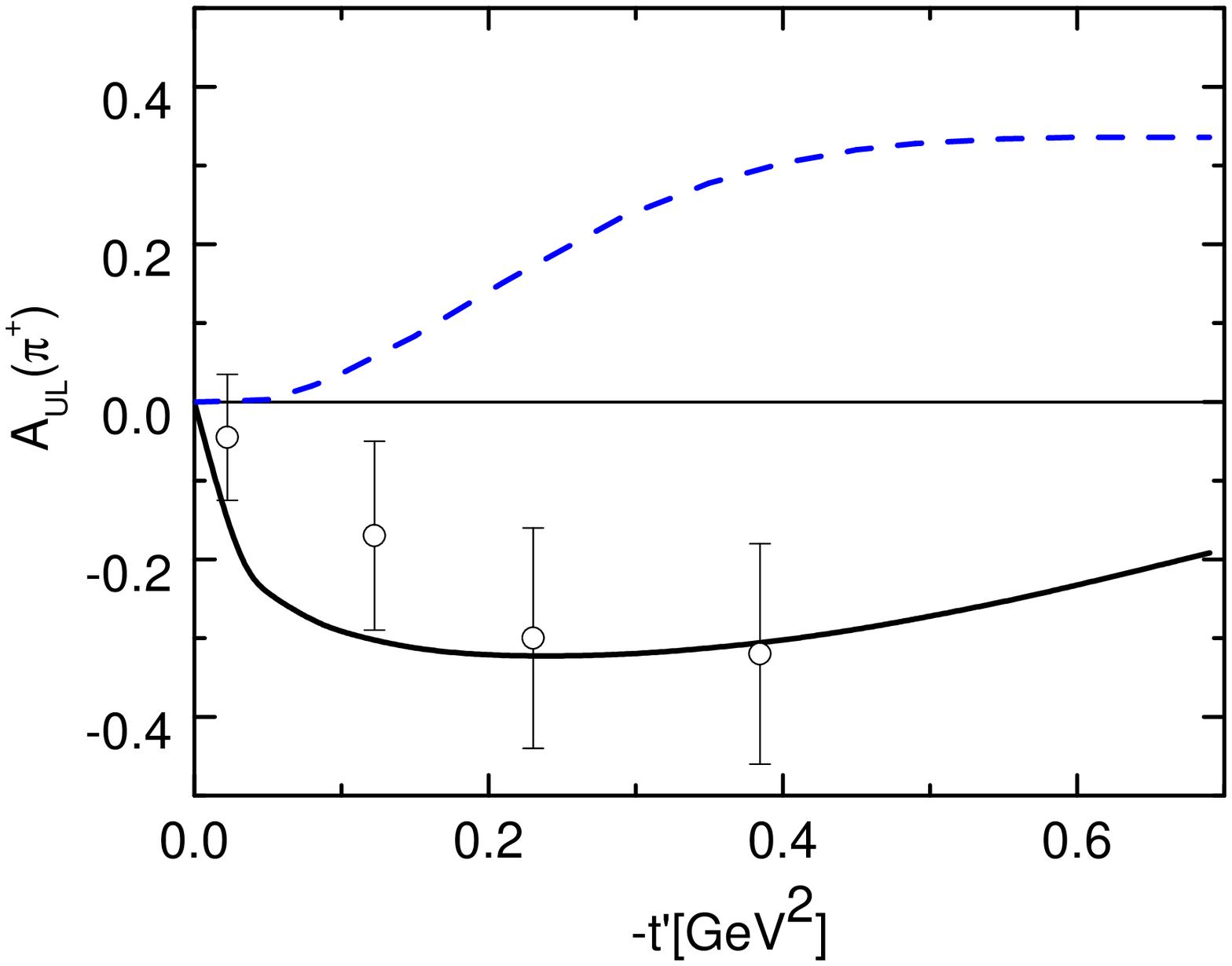}
  \caption{ \label{fig:5} Left: Predictions for the  $\sin{(\phi-\phi_s)}$
    moment at $Q^2=2.45\,\gev^2$  and $W=3.99\,\gev$ shown as solid
    lines~\protect\ci{GK5}. The dashed line represents the longitudinal 
    contribution to the $\sin{(\phi-\phi_s)}$ moment. Data are taken 
    from \protect\ci{Hristova}.}
  \caption{\label{fig:6} Right:  The asymmetry for a longitudinally polarized 
  target at $Q^2\simeq 2.4\,\gev^2$ and $W\simeq 4.1\,\gev$. The dashed line 
  is obtained disregarding the twist-3 contribution. Data are taken from 
  \protect\ci{hermes02}.}
  \end{center}
  \end{figure}

It is shown in Ref.\ \ci{GK5} that with the described model GPDs, the 
$\pi^+$ cross sections as measured by HERMES~\ci{HERMES07} are 
nicely fitted  as well as the transverse target asymmetries~\ci{Hristova}. This 
can be seen for $A_{UT}^{\sin{\phi_s}}$ from Fig.\ \ref{fig:1}. Also the 
$\sin(\phi-\phi_s)$ moment which is dominantly fed by an interference term 
of the the two amplitudes for longitudinally polarized photons (see Tab.\ 
\ref{tab:1}), is fairly well described as is obvious from Fig.\
\ref{fig:5}. Very interesting is also the asymmetry for a longitudinally 
polarized target which is dominated by the interference term between 
${\cal M}_{0-,++}$ which comprises the twist-3 effect, and the nucleon 
helicity-flip amplitude for $\gamma^*_L\to \pi$ transition, ${\cal M}_{0-,0+}$. 
Results for $A_{UL}^{\sin \phi}$ are displayed in Fig.\ \ref{fig:6} and compared 
to the data~\ci{hermes02}. Also in this case good agreement between theory 
and experiment is to be noticed. In both the cases, $A_{UT}^{\sin{\phi_s}}$ 
and $A_{UL}^{\sin \phi}$, the prominent role of the twist-3 mechanism is
clearly visible. Switching it off one obtains the dashed lines which are
significantly at variance with experiment. In this case the transverse 
amplitudes are only fed by the pion-pole contribution. The other transverse
target asymmetries quoted in Tab.\ \ref{tab:1} are predicted to be small 
in absolute value which is in agreement with experiment~\ci{Hristova}.
Thus, in summary, there is strong evidence for transversity in hard exclusive
pion electroproduction. It can be regarded as a non-trivial result that 
the transversity distributions determined from data on inclusive pion production 
lead to a transversity GPD which is nicely in agreement with target 
asymmetries measured in exclusive pion electroproduction. 

It is to be stressed that information on the amplitude ${\cal M}_{0-,++}$ can
also obtained from the asymmetries measured  with a longitudinally polarized
beam or with a longitudinally polarized beam and target. The first asymmetry, 
$A_{LU}^{\sin \phi}$, is dominated by the same interference term as $
A_{UL}^{\sin\phi}$ but diluted by the factor $\sqrt{(1-\varepsilon)/(1+\varepsilon)}$. 
Also the second asymmetry, $A_{LL}^{\cos \phi}$, is dominated 
by the interference term ${\cal M}_{0-,++}^*\,{\cal M}_{0-,0+}$. However, in
this case its real part occurs. For HERMES kinematics it is predicted to be
rather large and positive at small $-t^\prime$ and changes sign at 
$-t^\prime \simeq 0.4\,\gev^2$~\ci{GK5}. A measurement of these asymmetries
would constitute a serious check of the twist-3 effect.  
 
Although the main purpose of the work presented in Ref.\ \ci{GK5} is
focused on the analysis of the HERMES data one may also be interested in
comparing this approach with the Jefferson Lab data on the cross sections~\ci{horn06}. 
With the GPDs $\widetilde{H}, \widetilde{E}$ and $H_T$ in their present form
the agreement with these data is reasonable for the transverse cross section 
while the longitudinal one is somewhat too small. It is however to be stressed 
that the approach advocated for in Refs.\ \ci{GK1,GK5,GK4} is designed
for small skewness. At larger values of it the parameterizations of the GPDs 
are perhaps to simple and may require improvements. It is also important to 
realize that the GPDs are probed by the HERMES, COMPASS and HERA data only at 
$x$ less than about 0.6. One may therefore change the GPDs at large $x$ to 
some extent without changing the results for cross sections and asymmetries
in the kinematical region of small skewness. For Jefferson Lab kinematics, on 
the other hand, such changes of the GPDs may matter.

%%%%%%%%%%%%%%%%%%%%%%%%%%%%%%%%%%%%%%%%%%%%%%%%%%%%%%%%%%%%%%%%%%%%%%%%%%%%%%%%%%%%%%%%%%%%
\section{The GPD $E$}
%%%%%%%%%%%%%%%%%%%%%%%%%%%%%%%%%%%%%%%%%%%%%%%%%%%%%%%%%%%%%%%%%%%%%%%%%%%%%%%%%%%%%%%%%%%
In Ref.\ \ci{DFJK4} the electromagnetic form factors of the proton and
neutron have been utilized in order to determine the zero-skewness GPDs for 
valence quarks through the sum rules which for the case of the Pauli form
factor, reads
\be
F_2^{p(n)} = \int_{0}^{1} dx \Big[e_{u(d)}\,E^u_v(x,\xi=0,t) + e_{d(u)}\,E^d_v(x,\xi=0,t)\Big]\,.
\ee 
In order to determine the GPDs from the integral a parameterization of the GPD
is required for which the ansatz  
\be
E_v^a(x,0,t)= e_v^a(x) \exp\Big[t(\alpha^\prime_v \ln(1/x)+b_e^a)\Big]
\label{E-ansatz}
\ee
is made in a small $-t$ approximation~\ci{DFJK4}.  The forward limit of $E$ is
parameterized analogously to that of the usual parton distributions:
\be
e_v^a = N_a x^{\alpha_v(0)} (1-x)^{\beta_v^a}\,,
\label{ev-ansatz}
\ee
where $\alpha_v(0)$ ($\simeq 0.48$) is the intercept of a standard Regge
trajectory and $\alpha_v^\prime$ in Eq.\ \req{E-ansatz} its slope. The 
normalization $N_a$ is fixed from the moment 
\be
\kappa^a = \int dx E^a_v(x,\xi,t=0)\,,
\ee
where $\kappa^a$ is the contribution of flavor-$a$ quarks to the anomalous
magnetic moments of the proton and neutron ($\kappa^u=1.67$, $\kappa^d=-2.03$). 
A best fit to the data on the nucleon form factors provides the powers 
$\beta_v^u=4$ and $\beta_v^d=5.6$. However, other powers are not excluded 
in the 2004 analysis presented in \ci{DFJK4}; the most extreme set of powers,
still in agreement with the form factor data, is $\beta_v^u=10$ and $\beta_v^d=5$. 
The analysis performed in \ci{DFJK4} should be repeated since 
new form factor data are available from Jefferson Lab, e.g.\ $G_{E}^n$ and 
$G_M^n$ are now measured up to $Q^2= 3.5$ and $5.0\,\gev^2$,
respectively~\ci{GEn,GMn}. These new data seem to favor $\beta_v^u<\beta_v^d$. 
The zero-skewness GPDs $E_v$ are used as input to a double distribution from 
which the valence quark GPDs for non-zero skewness are constructed~\ci{GK4}.

In \ci{GK4}, following Diehl and Kugler~\ci{kugler},
$E$ for gluons and sea quarks has been estimated from positivity bounds and a
sum rule for the second moments of $E$ which follows from a combination of
Ji's sum rule~\ci{ji97} and the momentum sum rule of deep
inelastic lepton-nucleon scattering. It turned out that the valence quark 
contribution to that sum rule is very small, in particular if 
$\beta^u_v<\beta^d_v$, with the consequence of an almost exact cancellation 
of the gluon and sea quark moments. The GPDs $E^g$ and $E^{\rm sea}$ are 
parameterized analogously to $E_v$, see Eqs.\ \req{E-ansatz}, \req{ev-ansatz}. 
The normalization of $E^{\rm sea}$ is fixed by assuming that an appropriate 
positivity bound (see Refs.\ \ci{poby,burkardt}) is saturated while 
that of $E^g$ is determined from the sum rule. Several variants of $E$ have 
been exploited in \ci{GK4} in a calculation of $A_{UT}^{\sin{(\phi-\phi_s)}}$ 
within the handbag approach.  The results for a few variants are compared 
to the HERMES data on $\rho^0$ production~\ci{HERMES-rho} in 
Fig.\ \ref{fig:3}. Agreement between theory and experiment is to be noted. 
Similar agreement is obtained for the preliminary  COMPASS data~\ci{COMPASS-rho}. 
Combining both the experiments a negative value of 
$A_{UT}^{\sin{(\phi-\phi_s)}}$ for $\rho^0$ production is favored in agreement
with the theoretical results obtained in \ci{GK4}, only the 
extreme variant $\beta_v^u=10$ and $\beta_v^d=5$ (dashed-dotted line in Fig.\
\ref{fig:3}) seems to be ruled out. In \ci{GK4} predictions for 
$\omega$, $\rho^+$, $K^{*0}$ and $\phi$ productions are also given. Their 
comparison with forthcoming data from HERMES and COMPASS may lead to a fair 
determination of the GPD $E$.

With $E$ at hand one may exploit Ji's sum rule for the parton angular
momenta. At zero skewness the sum rule reads     
\be
\langle J^a\rangle = \frac12\big[ q^a_{20} + e^a_{20}\big]\,, \qquad
 \langle J^g\rangle = \frac12\big[ q^g_{20} + e^g_{20}\big]\,.
\ee
From a variant with $\beta_v^u=4$, $\beta_v^d=5.6$ and neglected $E^g$ 
and $E^{\rm sea}$ (solid line in Fig.\ \ref{fig:3}) for instance one obtains 
\be
\langle J^u \rangle = 0.250\,, \quad \langle J^d \rangle = 0.020\,, \quad 
\langle J^s \rangle = 0.015\,, \quad \langle J^g \rangle = 0.214\,, 
\ee
at the scale of $4\,\gev^2$. The angular momenta sum up to $\simeq 1/2$, the
spin of the proton. A very characteristic stable pattern is obtained in Ref.\ 
\ci{GK4}: For all variants investigated, $J^u$ and $J^g$ are large while 
the other two angular momenta are very small. The angular momenta of the 
valence quarks are $\langle J^u_v\rangle=0.222$ and $\langle
J^d_v\rangle=-0.015$. These values are identical to the results quoted in 
Ref.\ \ci{DFJK4} (for variant 1). They are in agreement with a recent 
lattice result~\ci{lattice}.

  %%%%%%%%%%%%%%%%%%%%%%%%%%%%%%%%%%%%%%%%%%%%%%%%%%%%%%%%%%%%%%%%%%%%%%%%%%%%
  \section{Summary}
  \label{sec:summary}
  %%%%%%%%%%%%%%%%%%%%%%%%%%%%%%%%%%%%%%%%%%%%%%%%%%%%%%%%%%%%%%%%%%%%%%%%%%%
Recent measurements of single spin asymmetries in hard meson
electroproduction has been reviewed. The data clearly show that a
leading-twist calculation of meson electroproduction within the handbag 
approach is insufficient. They demand higher-twist and/or power corrections 
which manifest themselves through substantial contributions from 
$\gamma_T^* \to V, P$ transitions.

A most striking effect is the target asymmetry $A_{UT}^{\sin \phi_s}$ in
$\pi^+$ electroproduction. The interpretation of this effect requires a large
contribution from the helicity non-flip amplitude ${\cal M}_{0-,++}$. Within 
the handbag approach such a contribution is generated by the helicity-flip or 
transversity GPDs in combination with a twist-3 pion wave function~\ci{GK5}.
This explanation establishes an interesting connection to transversity
parton distributions measured in inclusive processes. Further studies of 
transversity in exclusive reactions are certainly demanded. For instance, 
data on the asymmetries obtained with a longitudinally polarized beam and 
with likewise polarized beam and target would be very helpful in settling 
this dynamical issue.
Good data on $\pi^0$ electroproduction would also be highly welcome. They
would not only allow for an additional test of the twist-3 mechanism but also
give the opportunity to verify the model GPDs $\widetilde{H}$ and 
$\widetilde{E}$ as used in Ref.\ \ci{GK5}.
  
One may wonder whether the twist-3 mechanism does not apply to vector-meson 
electroproduction as well and offers an explanation of the experimentally 
observed $\gamma_T^*\to V_L$ transitions seen for instance in the SDME $r_{00}^{05}$.  
It however turned out that this effect is too small in comparison to the data.
The reason is that instead of the parameter 
$\mu_\pi$ the mass of the vector meson sets the scale of the twist-3 effect. 
This amounts to a reduction by about a factor of three. Further suppression 
comes from the unfavorable flavor combination of $H_T$ occurring for 
uncharged vector mesons, e.g.\ $e_u H_T^u-e_d H_T^d$ for $\rho^0$ production 
instead of $H_T^u-H_T^d$ for $\pi^+$ production. Perhaps the gluonic GPD 
$H_T^g$ may lead to a larger effect.
  
From the small value of the ratio of the longitudinal and transverse
electroproduction cross sections for $\rho^0$ and $\phi$ mesons it
also clear that the transitions from transversely polarized virtual photons
to likewise polarized vector mesons are large too. In the handbag
approach advocated in \ci{GK1} such transitions are also well
described. The infrared divergence occuring in collinear approximation 
is regularized by the quark transverse momenta in the modified perturbative approach.

  %%%%%%%%%%%%%%%%%%%%%%%%%%%%%%%%%%%%%%%%%%%%%%%%%%%%%%%%%%%%%%%%%%%%%%%%%%%%
  {\bf Acknowledgements} This work is supported  in part by the Heisenberg-Landau
  program and by the European Projekt Hadron Physics 2 IA in EU FP7. 

  \vskip 10mm 
  %%%%%%%%%%%%%%%%%%%%%%%%%%%%%%%%%%%%%%%%%%%%%%%%%%%%%%%%%%%%%%%%%%%%%%


\begin{thebibliography}{99}

\bibitem{rad96}A.~V.~Radyushkin,
  %``Asymmetric gluon distributions and hard diffractive electroproduction,''
  Phys.\ Lett.\  {\textbf B385} (1996)~333.
  %%CITATION = PHLTA,B385,333;%%

\bibitem{col96} J.C.\ Collins, L.\ Frankfurt and M.\ Strikman, 
  %``Factorization for hard exclusive electroproduction of mesons in QCD,''
  Phys.\ Rev.\ {\textbf D56}  (1997)~2982.
    %%CITATION = HEP-PH 9611433;%%

\bibitem{sterman} J.~Botts and G.~Sterman,
   Nucl.\ Phys.\ {\textbf B325} (1989)~62.
  
\bibitem{mul94}D.~Mueller {\it et al.},
    %``Wave functions, evolution equations and evolution kernels from light-ray
    %operators of {QCD},''
    Fortsch.\ Phys.\  {\textbf 42} (1994)~101.
    %%CITATION = HEP-PH 9812448;%% 

  \bibitem{rad98} A.~V.~Radyushkin,
    %``Symmetries and structure of skewed and double distributions,''
    Phys.\ Lett.\ {\textbf B449} (1999)~81.
    %%CITATION = HEP-PH 9810466;%%

  \bibitem{jakob} R.~Jakob and P.~Kroll,
  %``The Pion Form-Factor: Sudakov Suppressions And Intrinsic Transverse
    %Momentum,''
    Phys.\ Lett.\  {\textbf B315} (1993)~463
    [Erratum-ibid.\  {\textbf B319} (1993)~545].
     %%CITATION = PHLTA,B315,463;%%

  \bibitem{GK1} S.~V.~Goloskokov and P.~Kroll,
    %``Vector meson electroproduction at small Bjorken-x and generalized parton
    %distributions,''
   Eur.\ Phys.\ J.\ {\textbf C42} (2005)~281;
   {\it ibid.}  {\textbf C50} (2007)~829;
   {\it ibid.}  {\textbf C53} (2008)~367.

\bibitem{sapeta} M.~Diehl and S.~Sapeta,
    %``On the analysis of lepton scattering on longitudinally or transversely
    %polarized protons,''
    Eur.\ Phys.\ J.\  {\textbf C41} (2005)~515.
    %%CITATION = EPHJA,C41,515;%%

  \bibitem{Hristova} A.~Airapetian {\it et al.}  [HERMES Collaboration],
  %``Single-spin azimuthal asymmetry in exclusive electroproduction of pi+
  %mesons on transversely polarized protons,''
  arXiv:0907.2596 [hep-ex].
  %%CITATION = ARXIV:0907.2596;%%

\bibitem{HERMES-rho} A.~Airapetian {\it et al.}  [HERMES Collaboration],
  %``Exclusive rho-0 electroproduction on transversely polarized protons,''
  Phys.\ Lett.\  {\textbf B679} (2009)~100.
  %%CITATION = PHLTA,B679,100;%%

\bibitem{COMPASS-rho} G.\ Jegou [for the COMPASS collaboration],
to appear in Proceedings of DIS 2009, Madrid, Spain (2009)


\bibitem{GK4} S.~V.~Goloskokov and P.~Kroll,
  %``The target asymmetry in hard vector-meson electroproduction and parton
    %angular momenta,''
    Eur.\ Phys.\ J.\  {\textbf C59} (2009)~809.
    %%CITATION = EPHJA,C59,809;%%

\bibitem{GK5}  S.~V.~Goloskokov and P.~Kroll,
  %``An attempt to understand exclusive pi+ electroproduction,''
  arXiv:0906.0460 [hep-ph].
  %%CITATION = ARXIV:0906.0460;%%
 
\bibitem{horn06} H.~P.~Blok {\it et al.}  [Jefferson Lab Collaboration],
    %``Charged pion form factor between $Q^2$=0.60 and 2.45 GeV$^2$. I.
    %Measurements of the cross section for the ${^1}$H($e,e'\pi^+$)$n$ reaction,''
    Phys.\ Rev.\  {\textbf C78} (2008)~045202.
    %%CITATION = PHRVA,C78,045202;%%

\bibitem{bechler}   C.~Bechler and D.~Mueller,
  %``Generic modelling of non-perturbative quantities and a description of hard
  %exclusive $\pi^+$ electroproduction,''
  arXiv:0906.2571 [hep-ph].
  %%CITATION = ARXIV:0906.2571;%%

  \bibitem{diehl01} M.~Diehl,
  Eur.\ Phys.\ J.\ {\textbf C19} (2001)~485.

  \bibitem{hoodbhoy} P.~Hoodbhoy and X.~Ji,
  Phys.\ Rev.\ {\textbf D58} (1998)~054006.

\bibitem{passek} H.~W.~Huang, R.~Jakob, P.~Kroll and K.~Passek-Kumericki,
  %``Signatures of the handbag mechanism in wide-angle photoproduction of
  %pseudoscalar mesons,''
  Eur.\ Phys.\ J.\  {\textbf C33} (2004)~91.
  %%CITATION = EPHJA,C33,91;%%

\bibitem{huang}  H.~W.~Huang and P.~Kroll,
  %``Large momentum transfer electroproduction of mesons,''
  Eur.\ Phys.\ J.\  {\textbf C17} (2000)~423.
  %%CITATION = EPHJA,C17,423;%%


  \bibitem{braun90} V.~M.~Braun and I.~E.~Halperin,
  %``Conformal Invariance And Pion Wave Functions Of Nonleading Twist,''
  Z.\ Phys.\ {\textbf C48} (1990)~239.
  [Sov.\ J.\ Nucl.\ Phys.\  {\textbf 52} (1990\ YAFIA,52,199-213.1990)~126].
  %%CITATION = ZEPYA,C48,239;%%

  \bibitem{anselmino} M.~Anselmino, M.~Boglione, U.~D'Alesio, A.~Kotzinian,
    F.~Murgia, A.~Prokudin and C.~Turk,
    %``Transversity and Collins functions from SIDIS and e+ e- data,''
    Phys.\ Rev.\ {\textbf D75} (2007)~054032.
    %%CITATION = PHRVA,D75,054032;%%

  \bibitem{HERMES07} A.~Airapetian {\it et al.}  [HERMES Collaboration],
    %``Cross sections for hard exclusive electroproduction of pi+ mesons on a
    %hydrogen target,''
    Phys.\ Lett.\  {\textbf B659} (2008)~486.
    %%CITATION = PHLTA,B659,486;%%

\bibitem{hermes02} A.~Airapetian {\it et al} [HERMES Collaboration], 
      Phys.\ Lett.\ {\textbf B535} (2002)~85.

  \bibitem{DFJK4} M.~Diehl, T.~Feldmann, R.~Jakob and P.~Kroll,
    %``Generalized parton distributions from nucleon form factor data,''
    Eur.\ Phys.\ J.\  {\textbf C39} (2005)~1.
    %%CITATION = EPHJA,C39,1;%%




\bibitem{GEn} B.\ Wojtsekhowski {\it et al} [Jeffferson Lab E02-013 Collaboration],
in preparation; and http://hallaweb.jlab.org/experiment/E02-013/

\bibitem{GMn}J.~Lachniet {\it et al.}  [CLAS Collaboration],
  %``A Precise Measurement of the Neutron Magnetic Form Factor GMn in the
  %Few-GeV2 Region,''
  Phys.\ Rev.\ Lett.\  {\textbf 102} (2009)~192001.
  %%CITATION = PRLTA,102,192001;%%

  \bibitem{kugler}  M.~Diehl and W.~Kugler,
    %``Next-to-leading order corrections in exclusive meson production,''
    Eur.\ Phys.\ J.\  {\textbf C52} (2007)~933.
    %%CITATION = EPHJA,C52,933;%%

\bibitem{ji97} X.~D.~Ji,
  %``Gauge invariant decomposition of nucleon spin,''
  Phys.\ Rev.\ Lett.\ {\textbf 78} (1997)~610.
  %%CITATION = PRLTA,78,610;%%

\bibitem{poby} P.~V.~Pobylitsa,
  %``Positivity bounds on generalized parton distributions in impact  parameter
  %representation,''
  Phys.\ Rev.\  {\textbf D66} (2002)~094002.
  %%CITATION = PHRVA,D66,094002;%%

\bibitem{burkardt} M.~Burkardt,
  %``Some inequalities for the generalized parton distribution E(x,0,t),''
  Phys.\ Lett.\  {\textbf B582} (2004)~151.
  %%CITATION = PHLTA,B582,151;%%

\bibitem{lattice} Ph.\ H\"agler {\it et al} [LHPC collaboration],
arXiv:0705.4295 [hep-lat].

\end{thebibliography}
\end{document}